\documentclass[rsi,floatfix,
 amsmath,amssymb, reprint, twocolumn,author-year,
]{revtex4-1}

\usepackage{xr}
\usepackage{graphicx}
\usepackage{braket}
\usepackage{dcolumn}
\usepackage{bm}
\usepackage{color}
\usepackage[utf8]{inputenc}
\usepackage[T1]{fontenc}
\usepackage{mathptmx}
\usepackage{amsmath}
\usepackage{etoolbox}

\makeatletter
\def\@email#1#2{%
 \endgroup
 \patchcmd{\titleblock@produce}
  {\frontmatter@RRAPformat}
  {\frontmatter@RRAPformat{\produce@RRAP{*#1\href{mailto:#2}{#2}}}\frontmatter@RRAPformat}
  {}{}
}%
\makeatother
\begin{document}

\preprint{AIP/123-QED}

\title[Sample title]{A scanning resonator for probing quantum coherent devices}
\author{Jared Gibson}
\altaffiliation{These authors contributed equally.}
\affiliation{Department of Physics, University of Illinois at Urbana-Champaign, Urbana, IL 61801, USA}
\author{Zhanzhi Jiang}
\altaffiliation{These authors contributed equally.}
\affiliation{Materials Research Laboratory, University of Illinois at Urbana-Champaign, Urbana, IL 61801, USA}
\author{Angela Kou}
\affiliation{Department of Physics, University of Illinois at Urbana-Champaign, Urbana, IL 61801, USA}
\affiliation{Materials Research Laboratory, University of Illinois at Urbana-Champaign, Urbana, IL 61801, USA}
\affiliation{Holonyak Micro and Nanotechnology Lab, University of Illinois at Urbana-Champaign, Urbana, IL 61801, USA}

\date{\today}

\begin{abstract}
Superconducting resonators with high quality factors are extremely sensitive detectors of the complex impedance of materials and devices coupled to them. 
This capability has been used to measure losses in multiple different materials and, in the case of circuit quantum electrodynamics (circuit QED), has been used to measure the coherent evolution of multiple different types of qubits. 
Here, we report on the implementation of a scanning resonator for probing quantum coherent devices. 
Our scanning setup enables tunable coherent coupling to systems of interest without the need for fabricating on-chip superconducting resonators. 
We measure the internal quality factor of our resonator sensor in the single-photon regime to be $>10^4$ and demonstrate capacitive imaging using our sensor with zeptoFarad sensitivity and micron spatial resolution at milliKelvin temperatures. 
We then use our setup to characterize the energy spectrum and coherence times of multiple transmon qubits with no on-chip readout circuitry. 
Our work introduces a new tool for using circuit QED to measure existing and proposed qubit platforms.
\end{abstract}

\maketitle

\section{Introduction}
High-quality resonators, which are strongly affected by their impedance environment, can be used as sensitive detectors of the electric and magnetic response of different materials. 
A particularly useful example of this detection occurs when a microwave resonator is coherently coupled to a quantum coherent device. 
This interaction, referred to as circuit quantum electrodynamics (circuit QED), allows for measurement of the state of the quantum device by monitoring the resonator response \cite{blaisCircuitQuantumElectrodynamics2021, clerkHybridQuantumSystems2020}. 
Circuit QED has enabled high-fidelity readout of superconducting qubits \cite{wallraffStrongCouplingSingle2004, reedHighFidelityReadoutCircuit2010}, single-quanta detection of phonons \cite{satzingerQuantumControlSurface2018, oconnellQuantumGroundState2010, chuCreationControlMultiphonon2018, arrangoiz-arriolaResolvingEnergyLevels2019, slettenResolvingPhononFock2019} and magnons \cite{lachance-quirionResolvingQuantaCollective2017, lachance-quirionEntanglementbasedSingleshotDetection2020, xuQuantumControlSingle2023}, and ultra-high-sensitivity detection of electron spin resonances \cite{bienfaitControllingSpinRelaxation2016, sigillitoFastLowpowerManipulation2014, eichlerElectronSpinResonance2017, probstInductivedetectionElectronspinResonance2017}. 
In typical circuit QED setups, the resonator is fixed on the sample chip and the microwave fields of the resonator mode are partially shared with the quantum device \cite{blaisCircuitQuantumElectrodynamics2021, clerkHybridQuantumSystems2020}.

There are multiple scenarios, however, where it is beneficial to use a resonator that is not located on the same substrate as the quantum system we wish to probe. 
First, circuit QED has been proposed as a useful method for investigating the properties of quasiparticle excitations such as magnetic skyrmions \cite{psaroudakiSkyrmionQubitsNew2021, hirosawaMagnetoelectricCavityMagnonics2022} and Majorana zero modes \cite{mullerDetectionManipulationMajorana2013, yavilbergFermionParityMeasurement2015}. 
Such excitations are usually hosted in materials whose microwave loss precludes fabrication of the high-quality resonators necessary for circuit QED.
Second, for larger systems composed of multiple coupled quantum devices such as those used for quantum simulations and in quantum processors \cite{martinezFlatbandLocalizationInteractioninduced2023, rosenFlatbandDelocalizationEmulated2025, aruteQuantumSupremacyUsing2019}, space constraints make it difficult to couple an on-chip resonator to every device that we wish to measure.
Finally, for devices with large footprints where the spatial distribution of fields contains useful information, fixed on-chip resonators couple at a single location and it can be challenging to extract location-dependent properties of the device.

Scanning resonator probes are a flexible solution to the above-mentioned problems. 
A scanning resonator can be fabricated on a low-loss substrate, which could then be controllably coupled to the quantum system of interest.
Previous implementations of scanning microwave probes have mainly been focused on the low-quality factor regime and have been operated with photon numbers $>10^9$ \cite{laiModelingCharacterizationCantileverbased2008, jiangImplementingMicrowaveImpedance2023, caoMilliKelvinMicrowaveImpedance2023, shanJohnsonnoiselimitedCancellationfreeMicrowave2024, chuMicrowaveMicroscopyIts2020, barberMicrowaveImpedanceMicroscopy2022}. 
While these probes have provided invaluable information about the local electrodynamic response of multiple different materials \cite{chuMicrowaveMicroscopyIts2020, barberMicrowaveImpedanceMicroscopy2022}, the necessary usage of high powers for measurement make them unsuitable for probing the single-quanta energy spectrum and coherence properties of quantum devices.
Recently, a first step was taken toward dielectric detection in the low-power regime using a scanning resonator on a tip; a high-$Q$ resonator operating at low photon number was used to image the capacitive contrast between an aluminum capacitor and a silicon substrate \cite{degraafNearfieldScanningMicrowave2013,geaneyNearFieldScanningMicrowave2019}. 

\begin{figure*}[bt]
\includegraphics{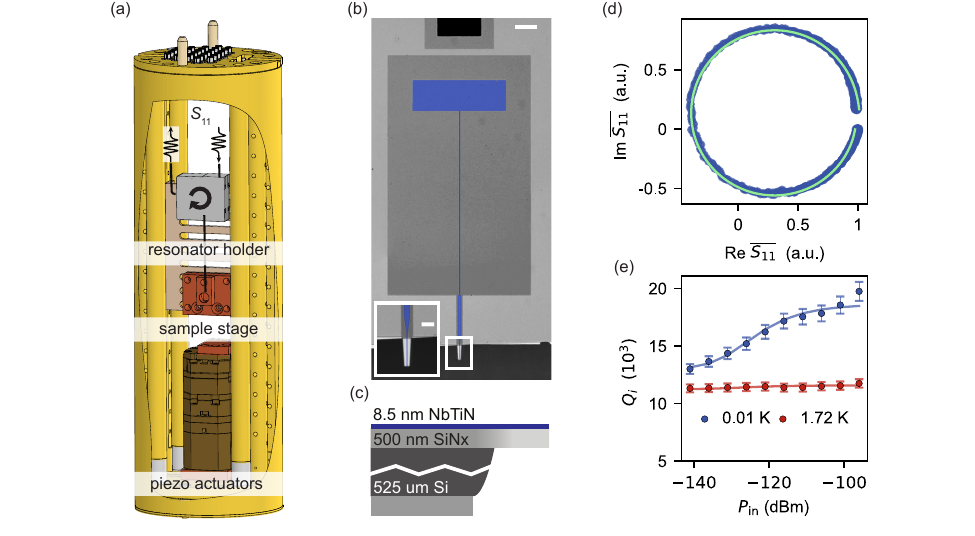}
\caption{\label{fig:setup} Scanning resonator setup. (a) Schematic of the sample puck for housing the scanning setup. (b) False-colored optical image of the lumped-element resonator sensor (scale bar 70 $\mathrm{\mu}$m)  and zoomed-in image of tip (inset; scale bar 15 $\mathrm{\mu}$m). (c) Materials stack used to create the resonator tip. (d) Normalized scattering parameter, $\overline{S_{11}}$, of the resonator. From the fit (solid line), we extract a resonance frequency of $f_0 = 7.955$ GHz and an internal quality factor $Q_i = 11900$ at 10 mK at an average photon number $\bar{n}< 1$. (e) Internal quality factor $Q_i$ at different mixing chamber temperatures as a function of input power $P_{\mathrm{in}}$. Solid lines are fits to a loss model including contributions from TLS and a power-independent loss.}
\end{figure*}

Here, we incorporate a high-quality superconducting resonator into a scanning setup. 
We use the resonator to perform capacitive imaging and then show its utility in measuring the time dynamics of a quantum coherent system by coupling it to a transmon qubit. 
Our resonator design is simple, composed of a lumped-element resonator terminating in a narrow tip on a cantilever, which allows our scanning resonator to have a spatial resolution of $\sim2~\mathrm{\mu}$m. 
We measure the resonator internal quality factor to be $Q_i>10^4$ in the single-photon regime and up to temperatures $>1$~K. 
The capacitive sensitivity of our resonator is comparable to state-of-the-art high-power scanning microwave probes \cite{chuMicrowaveMicroscopyIts2020, barberMicrowaveImpedanceMicroscopy2022, shanAdvancedMicrowaveImpedance2024} with the benefit of operation in the single-photon regime. 
We benchmark our resonator as a probe of coherent quantum devices by using our tip to measure the energy spectrum and coherence properties of multiple transmon qubits located on a single chip with no on-chip readout circuitry. 
Our scanning resonator system thus opens avenues for probing the dielectric response and the time dynamics of a multitude of quantum systems.

\section{Experimental Setup}  
\subsection{Scanning Platform}
The scanning platform is installed in a cryogen-free dilution refrigerator (ProteoxMX, Oxford Instruments) with a base temperature of 10 mK. 
As shown in Fig. \ref{fig:setup}(a), the resonator holder, sample stage, a scanner (ANSxyz100, AttoCube Systems AG), and a set of piezo positioners (ANPz102 and ANPx101, AttoCube Systems AG) are installed in a sample puck. 
The entire scanning platform is enclosed within an aluminum shield lined with Eccosorb foam.
Housing the scanning platform in a sample puck allows us to use the fast-exchange capabilities of our dilution refrigerator for rapid sample characterization; the exchange cycle using a sample puck is $<12$~hours. 
At the base temperature, the maximum single-frame scan window is 45 $\mathrm{\mu}$m $\times$ 45 $\mathrm{\mu}$m, and the complete positioning range is 5 mm $\times$ 5 mm.
The resonator is held fixed while the sample stage is moved by the piezo actuators.
We connect the sample stage to the puck with a flexible copper braid to thermalize the sample. In this configuration, the base temperature remains below 25 mK when the scanning speed is 500 nm/s.

\begin{figure*}[bt]
\includegraphics{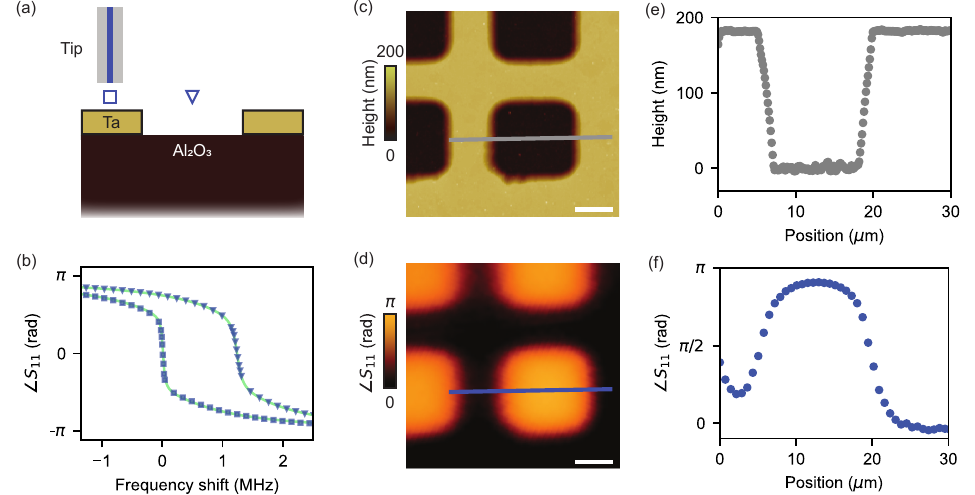}
\caption{\label{fig:imaging}Capacitive imaging capabilities. (a) Schematic of tip (blue) and cantilever (gray) over sample consisting of a patterned Ta thin film on an Al$_2$O$_3$ substrate. Square and triangle markers indicate tip position during measurement of sensor responses shown in (b). (b) Sensor response over Ta (squares) and over Al$_2$O$_3$ (triangles). $f_0$ obtained from fit (solid lines) indicates higher resonance frequency due to smaller $C_{\mathrm{ts}}$ over Al$_2$O$_3$. (c) ((d)) Atomic force microscopy (resonator phase) scan of sample region chosen for capacitive imaging; scale bars are 5 $\mathrm{\mu}$m. Gray (blue) line indicates position of the linecut plotted in (e) ((f)). (e) ((f)) Linecut of atomic force microscopy (resonator phase) scan over a 10 $\mu$m $\times$ 10 $\mu$m hole in 180 nm thick Ta film. We observe good agreement between the resonator measurements and measured sample topography. The tip is kept at 500 nm above the Ta surface when taking the data in (b), (d) and (f).}
\end{figure*}

\subsection{Sensor Implementation}
The goals of our experiment are to use our resonator as a sensitive probe of capacitance and as a readout resonator for probing quantum coherent systems.
For strong coupling to a sample, the coupling rate (controlled by the ratio of the tip-sample coupling capacitance $C_{\mathrm{ts}}$ to the resonator capacitance $C$) needs to be large compared to the linewidth of the resonator $\kappa = \omega_r/Q$ and the loss of the sample.
We additionally desire to ensure that the resonator imparts minimal additional losses to the probed quantum system.
We have thus optimized our resonator design to minimize its native capacitance and chosen a materials stack to maximize its quality factor.
As shown in Fig. \ref{fig:setup}(b) and (c), the resonator was fabricated using an 8.5-nm thick NbTiN film deposited on a $\mathrm{SiN_{x}/Si/SiN_{x}}$ substrate. 
The high kinetic inductance of NbTiN (68 pH/sq) allows us to minimize the resonator capacitance while still maintaining the resonator frequency in the range of readily-available microwave components (4-12 GHz).
A 2-$\mathrm{\mu}$m wide wire of length 550 $\mathrm{\mu}$m  extending from the pad acts as the resonator inductance with $L \approx$ 19 nH. 
The blue pad in Fig. \ref{fig:setup}(b) contributes to the resonator inductance and controls the coupling of the resonator to an external microwave feedline that we use to send probe microwaves.
The total resonator capacitance is $\approx$ 21 fF for our resonator frequency of 7.955 GHz.

We chose the SiN$_x$/Si/SiN$_x$ substrate since these substrates have been shown to support resonators with internal quality factors $Q_i \geq 10^4$ at low photon powers \cite{vissersLowLossSuperconducting2010, martinisDecoherenceJosephsonQubits2005, chistoliniPerformanceSuperconductingResonators2024}. 
Moreover, the selectivity of etching between Si and SiN$_x$ allows us to use standard fabrication methods to make the cantilever for our resonator tip. 
We defined our resonator sensor using a combination of dry etching and wet etching techniques; fabrication details are provided in the supplement.

Our sensor was also designed to provide local information about the sample. As shown in the inset of Fig. \ref{fig:setup}(b), we terminate the inductor at a narrow $2$-$\mu$m tip on a cantilever of thickness 500 nm. The cantilever structure allows us to safely bring the tip near the surface so that we can take advantage of the spatial resolution afforded by the narrow tip.

\subsection{Resonator Measurements}
We characterized our resonator in the scanning platform using microwave spectroscopy as shown in Fig. \ref{fig:setup}(a). 
We found the internal quality factor of our resonator by measuring the complex scattering parameter $S_{11}$ reflected off of the feedline coupled to our resonator. 
We plot the normalized scattering parameter $\overline{S_{11}}$ in Fig. \ref{fig:setup}(d) and perform a fit (solid line) to extract a resonance frequency $f_0 = 7.955$ GHz and an internal quality factor $Q_i = 11900$ at 10 mK using an input power of -141 dBm, which corresponds to $\bar{n}<1$ in the resonator.

To understand the loss mechanisms of our resonator, we measured the $Q_i$ at different temperatures and input powers. 
At temperatures below 1 K, the resonator $Q_i$ improves for increasing input powers as shown in Fig. \ref{fig:setup}(e), suggesting saturation of two-level systems (TLS) present in the materials stack in Fig. \ref{fig:setup}(c). 
This power-dependent $Q_i$ is consistent with observations of TLS loss in superconducting resonators on SiN$_x$/Si substrates \cite{vissersLowLossSuperconducting2010, chistoliniPerformanceSuperconductingResonators2024}.
The data fits well to a loss model (solid lines) including contributions from a power-independent loss and TLS losses. 
The power-independent losses may be due to pair-breaking radiation or radiative losses to electromagnetic modes hosted by the sample puck (see supplement for more details). 
We note that the resonator maintains a $Q_i >10^4$ above 1 K. 
At these temperatures, the $Q_i$ depends only very weakly on input power, indicating that TLS's are no longer a major contributor to microwave losses. 
We attribute losses at higher temperatures to thermally-excited quasiparticles in the superconducting film. 
We also observed a temperature-dependent frequency shift of the resonator consistent with thermally-excited quasiparticles at high temperatures (see supplement for details).

\section{Capacitive Imaging}
We next investigated the capacitive sensitivity of our sensor using a sample composed of patterned Ta on a sapphire (Al$_2$O$_3$) substrate. 
When we positioned the tip of our resonator over the Ta film as in Fig. \ref{fig:imaging}(a) and (b) (squares), we measured a larger $C_{\mathrm{ts}}$ than when the tip is over Al$_2$O$_3$ (triangles). 
This change in $C_{\mathrm{ts}}$ causes the resonator frequency to shift upward, as shown in Fig. \ref{fig:imaging}(b). 
When $C_{\mathrm{ts}}$ is small compared to the resonator self-capacitance, the shift of $f_0$ is approximately proportional to the change in $C_{\mathrm{ts}}$ (see supplementary information), allowing us to estimate the capacitive sensitivity of our instrument using    
\begin{align}
\delta C_{\mathrm{ts}} = \delta f \frac{\partial{C_{\mathrm{ts}}}}{\partial{f}} \\
\approx \alpha \delta f,  
\end{align} 
where $\delta f$ is the spectral density of the frequency noise and $\alpha$ is a proportionality constant that relates changes in $f_0$ to changes in $C_{\mathrm{ts}}$. 
At single photon powers, the sensitivity at a typical measurement bandwidth of 1 Hz is 3 zF/Hz$^{-1/2}$ when the tip-sample distance is 500 nm. 
This capacitive sensitivity is comparable to other microwave microscopy techniques but with a significantly lower applied power \cite{laiNanoscaleMicrowaveMicroscopy2011,shanJohnsonnoiselimitedCancellationfreeMicrowave2024,degraafNearfieldScanningMicrowave2013}.
Operating in this low-power regime while maintaining a high sensitivity allows us to probe quantum coherent devices via the tip sample capacitance without negatively perturbing their quantum states. 
The sensitivity is mainly limited by the mechanical vibrations in the dilution refrigerator \cite{barberCharacterizationTwoFastTurnaround2024}.

We estimated the spatial resolution of our sensor by imaging etched squares in the Ta film that have been previously characterized using atomic force microscopy (AFM) (Fig. \ref{fig:imaging}(c)).
We fixed the phase offset of our resonator response such that we measure zero phase in the resonator response when the tip is over Ta. 
As we scanned our tip over the sample, we observed changes in phase corresponding to changes in $C_{\mathrm{ts}}$ as shown in Fig. \ref{fig:imaging}(d). 
We find qualitative agreement between the capacitive image and the topography of the sample as measured in AFM. 
Linecuts of the AFM and resonator data across a Ta/$\mathrm{Al_2O_3}$ boundary are shown in Fig. \ref{fig:imaging}(e) and (f).
We note that the resonator signal transition at the boundary is $\approx$ 2 $\mathrm{\mu}$m broader than the width of the physical boundary measured by AFM, demonstrating a spatial resolution matching the tip width. 
The drive power (-135 dBm) used to obtain the image in Fig. \ref{fig:imaging}(d) corresponds to about $\bar{n} \approx 1$ in the resonator, which is beneficial for probing local properties of quantum coherent devices.   

\section{Transmon Characterization}
\begin{figure}
\includegraphics{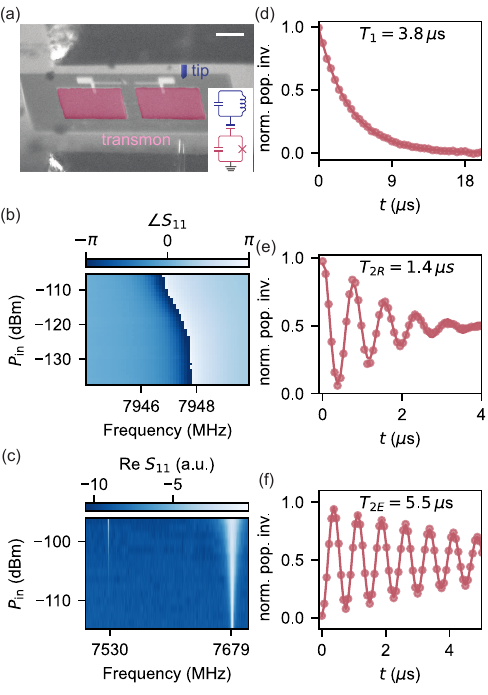}
\caption{\label{fig:single transmon} Single-transmon characterization. (a) False-colored optical image of the resonator tip (blue) and the transmon (pink); scale bar is 40 $\mathrm{\mu}$m. (b) Resonator spectroscopy at increasing microwave drive powers. The phase offset is set to make the phase zero when the resonator is off-resonance. (c) Transmon spectrum measured using two-tone spectroscopy of the resonator-transmon system. Measured transmon $T_1$, $T_{2R}$, and $T_{2E}$ times shown in (d)-(f). The tip-sample distance is 2.5 $\mu$m for these measurements.}
\end{figure}

In addition to capacitive sensing, we aim to use our resonator to measure the time dynamics of quantum coherent systems.
We demonstrate this capability by using our resonator to characterize a transmon \cite{schreierSuppressingChargeNoise2008}. 
Figure \ref{fig:single transmon}(a) shows an optical image of the tip (blue) positioned over a transmon (pink). 
The coupling between the resonator and the transmon is controlled by the tip-sample capacitance and can be changed by varying the distance between the tip and the sample. 

\begin{figure}
\includegraphics{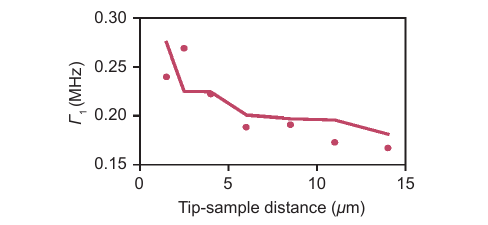}
\caption{\label{fig:transmon loss} Relaxation rate of the transmon as a function of the tip-sample distance. The dots are the transmon relaxation rate measured at each tip-sample distance. The solid line is the sum of the simulated Purcell loss rate and a tip-sample distance-independent loss rate of 0.1 MHz.}
\end{figure}

We first investigated the resonator response in the presence of a transmon.
We operated in the dispersive regime for the transmon-resonator system, where the transmon induces a state-dependent frequency shift on the resonator \cite{blaisCavityQuantumElectrodynamics2004}. 
When probing the resonator at high powers, we expect the resonance to be located at its bare-frequency, but as the drive power is decreased to the few-photon regime, the resonance will shift to a dressed frequency due to its coupling to the transmon \cite{reedHighFidelityReadoutCircuit2010, bishopResponseStronglyDriven2010, boissonneaultImprovedSuperconductingQubit2010}.
We indeed observe this expected non-linear behavior in the resonator response as a function of readout power as shown in Fig. \ref{fig:single transmon}(b). 
  
We then used the resonator to perform two-tone spectroscopy to measure the transmon energy spectrum as shown in Fig. \ref{fig:single transmon}(c). 
Here, we applied two microwave tones to the resonator: one to probe the resonator response at a fixed frequency and a second to excite the transmon from its ground state to higher energy levels through the resonator-transmon coupling \cite{blaisQuantuminformationProcessingCircuit2007}. 
By sweeping the frequency and power of the second microwave tone, we observed two transitions that appear at 7528 MHz and 7679 MHz. 
The higher-frequency transition corresponds to the ground $\ket{g}$ to first excited state $\ket{e}$ transition. 
The lower frequency transition is only excited at higher powers; this transition corresponds to the two-photon transition that excites $\ket{g}$ to the second excited state $\ket{f}$.

Next, we characterized the coherence properties of the transmon, including the relaxation time $T_1$, the Ramsey dephasing time $T_{2R}$, and the echo dephasing time $T_{2E}$. 
To measure $T_1$, we prepared the transmon in the excited state $\ket{e}$ with a $\pi$ pulse and allow it to evolve for a time $t$, after which the qubit state was measured using the resonator. Figure \ref{fig:single transmon}(d) shows the results of a qubit relaxation measurement. 
We find a relaxation time of $T_1 = 3.8$ $\mathrm{\mu}$s. 
For the Ramsey measurement, we initialized the transmon in the superposition state $1/2(\ket{g} + \ket{e})$ with a $\pi/2$ pulse, waited for a time $t$, applied a second $\pi/2$ pulse and then read out the transmon state. 
We find the Ramsey dephasing time to be $T_{2R} = 1.4$ $\mu s$ (Fig. \ref{fig:single transmon}(e)), which indicates that the qubit coherence is affect by pure dephasing processes. 
In order to identify the contribution of low-frequency noise, we performed a Hahn echo experiment where a $\pi$ pulse is sandwiched in the middle of a Ramsey pulse sequence. We observe a significant increase in the $T_{2E}$ time as shown in Fig.~\ref{fig:single transmon}(f).
The low-frequency dephasing processes are likely due to the slow mechanical vibrations of the dilution refrigerator.

To investigate the source of qubit relaxation in our setup, we measured $T_1$ at various tip-sample distances and calculated the corresponding relaxation rate $\mathnormal{\Gamma}_1 = 1 / T_1$, as shown in Fig. \ref{fig:transmon loss}. The increase of $\mathnormal{\Gamma}_1$ with the decreasing tip-sample distance is consistent with Purcell loss \cite{purcellResonanceAbsorptionNuclear1946}, since the qubit-resonator coupling becomes stronger at smaller tip-sample distances (see details in the supplement). 
The measured loss is well-captured by a model including Purcell loss and a tip-sample distance-independent loss.
Importantly, we do not find additional tip-sample capacitance-dependent contributions to the transmon loss beyond the predicted Purcell loss, which suggests that the scanning resonator does not induce losses to the transmon beyond the Purcell effect.
This Purcell loss can be mitigated by adding Purcell filters to the resonator tip in future implementations \cite{jeffreyFastAccurateState2014}.

\section{Multi-transmon Measurements}

\begin{figure}[t]
\includegraphics{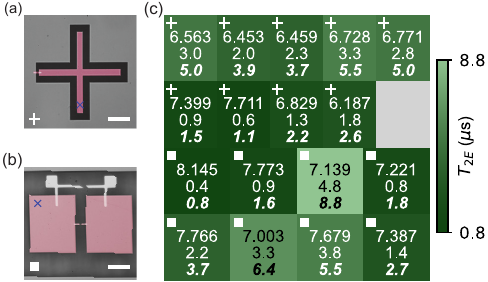}
\caption{\label{fig:multi transmon}  Multi-qubit characterization. (a) False-colored optical image of a measured Xmon qubit (scale bar 60 $\mu$m). (b) False-colored optical image of a planar-capacitor transmon qubit (scale bar 40 $\mu$m). Blue markers in (a) and (b) indicate the tip position during transmon measurements. (c) Chip mapping of the $\ket{g}$-to-$\ket{e}$ transition frequencies $f_{ge}$ (top, GHz), relaxation time $T_1$ (middle, $\mu$s) and echo dephasing time $T_{2E}$ (bottom, $\mu$s). Tile color corresponds to $T_{2E}$. Markers in upper left corners of tiles denote whether the transmon is of the Xmon (plus marker) or planar-capacitor (square marker) type.}
\end{figure}

The scanning capability of our sensor allows us to probe the response of multiple devices with various geometries without fabricating readout circuitry on the sample chip. 
This capability is particularly useful for high-throughput characterization of quantum devices that are either difficult to access or where readout circuitry may be located on separate chip layers \cite{rosenberg3DIntegratedSuperconducting2017, holman3DIntegrationMeasurement2021}.
Here we measured an array of 17 separate transmon qubits with no on-chip readout resonators. Nine of the transmons have an Xmon geometry (Fig. \ref{fig:multi transmon}(a)) \cite{barendsCoherentJosephsonQubit2013} and the other eight have a planar-capacitor geometry (Fig. \ref{fig:multi transmon}(b)) \cite{paikObservationHighCoherence2011}.

For characterizing each individual transmon, we moved the transmon under the resonator tip using the positioner. The blue markers in Fig. \ref{fig:multi transmon}(a) and (b) indicate the tip location relative to the transmon during measurement. 
The tip-sample distance is kept at 2.5 $\mathrm{\mu}$m for each measurement. Figure \ref{fig:multi transmon}(c) shows the $f_{ge}$, $T_1$, and $T_{2E}$ of each transmon, with each tile representing the location of a transmon. 
The characteristic energies and geometry of each transmon are provided in the supplement.

\section{Conclusions and Outlook}

We have demonstrated that our scanning resonator system is a useful tool for characterizing quantum coherent devices. 
Using a high-quality factor superconducting resonator as our sensor and minimizing the native capacitance of the sensor, we achieved capacitive sensitivity down to several zF/Hz$^{1/2}$ at single-photon powers. 
Furthermore, we showed that our sensor is capable of coherent interactions with quantum coherent devices by measuring the lifetime and coherence time of a transmon with no on-chip circuitry. 
We found that the transmon loss was consistent with contributions independent of the tip-sample capacitance and Purcell loss due to the resonator, which indicates that the tip itself did not contribute additional loss to the transmon.
Improvements in shielding in our setup and the addition of a Purcell filter on our tip could mitigate such losses in future devices. 
Finally, we used the scanning capability of our system to characterize multiple transmons on a single chip, which demonstrates its utility as a characterization tool for quantum processing architectures where readout and qubit circuitry are located on separate layers.

We have also identified possible improvements that could be made to our scanning setup. In the few-photon regime, the resonator quality factor is limited by TLS and power-independent loss channels which may be related to photons decaying into spurious modes and non-equilibrium quasi-particles induced by external radiations. 
Future optimization of tip fabrication, packaging, and shielding can further improve the resonator quality factor. 
Additionally, the capacitive sensitivity of the setup is limited by the vibrational noise of the fridge, which can be mitigated by floating the scanning platform with a spring stage \cite{quaglioSubKelvinScanningProbe2012}.

We expect that our scanning resonator will be useful for probing multiple different materials systems.
For instance, our scanning resonator can be used to investigate collective excitations in mesoscopic condensed matter systems at the single-quantum level \cite{clerkHybridQuantumSystems2020, molerImagingQuantumMaterials2017}. 
It can also be used to perform imaging of local loss and decoherence sources in solid-state qubits and perform readout of qubits incorporated into arrays for simulation of many-body physics \cite{martinezFlatbandLocalizationInteractioninduced2023, rosenFlatbandDelocalizationEmulated2025}.
Our scanning resonator system thus opens multiple avenues for new experiments in materials science, quantum information, and quantum simulation.
\section*{Supplementary Material}
The details about resonator fabrication, microwave measurement, calibration of tip-sample distance and capacitive sensitivity, resonator and transmon loss analysis, and information of the transmon chip are given in the supplementary material.
\begin{acknowledgments}
We thank M. Barber, L. Bishop-van Horn, B. E. Feldman, Z-X. Shen, and K. Moler for advice on the scanning setup, and G. Mensing for helpful discussions about resonator fabrication. We are also grateful to R. S. Goncalves, K. Nie, A. Bista, A. Baptista for assistance with transmon fabrication, and X. Cao, K. Singirikonda, P. Kim, C. Purmessur for assistance with qubit measurement and data analysis. The building of the experimental setup and tip fabrication was funded by the Center for Quantum
Sensing and Quantum Materials, an Energy Frontier Research Center funded by the U. S. Department of Energy,
Office of Science, Basic Energy Sciences under Award
DE-SC0021238. This work was partially supported by the NSF Quantum Leap Challenge Institute for Hybrid Quantum Architectures and Networks (NSF Award 2016136), a Multidisciplinary University Research Initiative of the Office of Naval Research (ONR) Award No. N00014-22-1-2764 P0000, and the Air Force Office of Scientific Research under award number FA9550-23-1-0690. Z. Jiang is partially supported by the Illinois Quantum Information Science and Technology Center (IQUIST) Postdoctoral Fellowship.
The resonator and transmons were fabricated using the factilities at the Materials Research Laboratory Central Research Facilities, the Micro-Nano-Mechanical Systems Cleanroom Laboratory, and the Holonyak Micro \& Nanotechnology Lab.
\end{acknowledgments}

\section*{Data Availability Statement}
The data that support the findings of this study are openly available in the Illinois Data Bank at databank.illinois.edu, reference number [reference number].

\bibliography{main}

\end{document}


\preprint{AIP/123-QED}

\renewcommand{\thefigure}{S\arabic{figure}}
\renewcommand{\thetable}{S\Roman{table}}
\newcommand{\Tabref}[1]{Table~\ref{#1}} 
\newcommand{\minus}{\scalebox{0.75}[1.0]{$-$}}

\title{Supplementary information for: A scanning resonator for probing quantum coherent devices}

\maketitle

\section{Resonator Fabrication}
Our resonator chip was fabricated from a Si substrate with capping layers of 500 nm SiN$_x$ on both sides. 
A blanket film of 8.5 nm NbTiN was deposited on the top side of the wafer. 
We first etched the bottom SiN$_x$ using optical lithography and dry etching with CHF$_3$ to define locations where the Si will be removed to form the cantilever and tip frame.
The coupon was then sealed in a custom sample holder from AMMT, which prevented the device side from being exposed to the wet etch solution.
The chip and the holder were then placed in a 30 \% KOH solution kept at 63 degrees Celsius using a hot plate and teflon stir bar at 180 rpm for 24 hrs, which etched the 525 $\mathrm{\mu}$m Si substrate. 
The top layer of SiN$_x$ acts as an etch stop so that the KOH does not affect the NbTiN on the device side of the coupon.
Next, we patterned the resonator and feedline in the NbTiN using optical lithography and dry etching with SF$_6$ and O$_2$. 
Finally, we defined the frame and cantilever on the top SiN$_x$ using an aluminum etch mask and dry etching with CHF$_3$.
The tips were then mechanically released in a solution of IPA to limit the Si residue on the resonator surface. 
The chip was wirebonded into a printed circuit board before being mounted in the sample puck.
 
\section{Microwave Measurement}
\begin{figure}[h]
\includegraphics{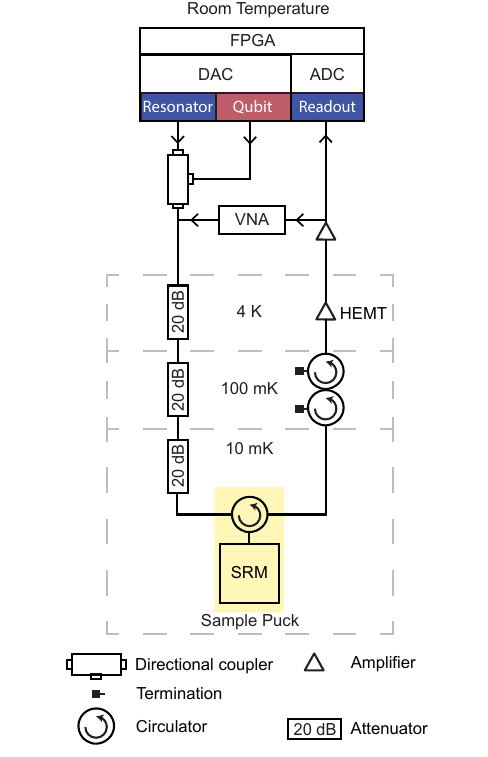}
\caption{\label{fig:wiring} Microwave electronics and fridge wiring diagram for experiments described in the main text. Resonator characterization and capacitive imaging experiments are done using a vector network analyzer (VNA). Transmon characterization experiments are performed using an FPGA.}
\end{figure}

For resonator characterization and capacitive imaging experiments described in the main text, we used a Keysight P9374B vector network analyzer (VNA) as shown in Fig. \ref{fig:wiring}. 
The output of the VNA is connected to the input line of the dilution refrigerator and the signal is attenuated at the 4 K, 100 mK, and 10 mK plates before reflecting off the scanning resonator microscope (SRM) inside the sample puck. 
The reflected signal is directed toward the output line by a circulator with a cryogenic isolator to prevent unwanted signals from reaching the resonator or sample via the output line. 
The signal is then amplified using a low-noise HEMT on the 4 K plate and a set of room-temperature amplifiers before reaching the input port of the VNA. 
For transmon characterization, we used a Quantum Instrumentation Control Kit (QICK) based on FPGA-controlled RF ADC/DAC channels \cite{stefanazziQICKQuantumInstrumentation2022}.
We used separate DAC channels for resonator and qubit pulse generation, which were connected to the input line via a directional coupler. 
We processed the reflected signal using a single ADC channel on the QICK.   

\begin{figure*}
\includegraphics{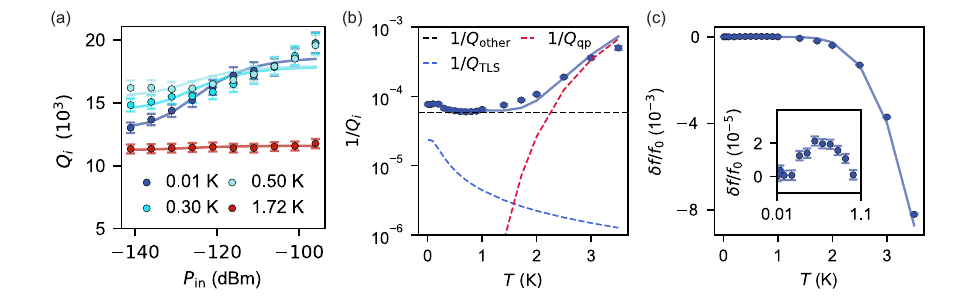}
\caption{\label{fig:resonator loss} Resonator loss mechanisms. (a) Additional temperature data showing an initial increase in $Q_i$ due to thermal saturation of the TLS bath, followed by a decrease at higher temperatures due to quasiparticles. Solid lines are fits to a TLS loss model. (b) Resonator loss $1/Q_i$ plotted for $P_{\mathrm{in}} = -141$ dBm at different temperatures. Solid line is a fit including losses from TLS (contribution in blue dashed line), quasiparticles (contribution in red dashed line), and a temperature-independent loss (contribution in black dashed line).  (c) Fractional frequency shift of the resonator for the same input power and temperatures as in panel (b), showing a small positive frequency shift at low temperatures (inset) giving way to a negative frequency shift. Solid line is a fit to a model accounting for the frequency shift due to quasiparticles.}
\end{figure*}
\section{Resonator Response Fitting and Photon Number Extraction}
We extracted the internal and coupling quality factors ($Q_i$ and $Q_c$, respectively) and resonant frequency $f_0$ of our resonator by fitting reflection data to the equation 
\begin{align}
\label{eq: reflection}
S_{11} = B e^{i\tau f} \left( 1 - \frac{2 \frac{Q}{Q_c}e^{i \theta}}{1 + 2iQ\frac{f - f_0}{f_0}}\right),
\end{align}
where $B$ is a complex number that accounts for the attenuation and a constant phase offset, the frequency-dependent phase factor $e^{i\tau f}$ addresses the electrical delay, $1/Q = 1/Q_i + \cos(\theta)/Q_c$ is the total quality factor, $\theta$ is an asymmetry factor introduced for impedance mismatch \cite{altoeLocalizationMitigationLoss2022}. 
The solid line in Fig. 1(d) in the main text is an example of this fitting procedure, where the asymmetry parameter is $\theta = -0.16$ rad. 
After extracting the fit parameters, we computed the average photon number using
\begin{align}
\label{eq: reflection}
\bar{n} = \frac{2 Q^2 P_{\mathrm{in}}}{\pi Q_c h f_0^2},
\end{align}
where $h$ is Planck's constant \cite{altoeLocalizationMitigationLoss2022}. 

\section{Resonator Loss Analysis}
We investigated the resonator loss mechanisms by fitting the input power ($P_{\mathrm{in}}$) and temperature ($T$) dependence of the internal quality factor $Q_i$ to model functions of multiple loss channels. The power dependence of $Q_i$ is mainly related to the two-level systems (TLS).
At T < 1 K, TLS get saturated as $P_{\mathrm{in}}$ approaches a critical power $P_c$, making the TLS loss decrease as
\begin{align}
\delta_{\mathrm{TLS}} = \frac{\delta^0_{\mathrm{TLS}}\tanh(\hbar\omega/2k_BT)}{\sqrt{1 + P_{\mathrm{in}}/P_c}},
\end{align}
where $\delta^0_{\mathrm{TLS}}$ is the TLS loss in the low-power low-temperature limit \cite{mcraeMaterialsLossMeasurements2020}. The power-dependence of $Q_i$ can therefore be fitted to 
\begin{align}
1/Q_i = 1/Q_{\mathrm{sat}} + 1/Q_{\mathrm{TLS}},
\end{align}
where $Q_{\mathrm{sat}}$ is the resonator internal quality factor when all TLS are saturated in the high-power limit, and $Q_{\mathrm{TLS}} = 1/\delta_{\mathrm{TLS}}$. We extract $\delta^0_{\mathrm{TLS}} = 2.3e-5$ and $Q_{\mathrm{sat}} \sim 18000$ from the fitting shown in Fig. 1(e) and Fig. \ref{fig:resonator loss}(a).

In the low-power limit, TLS loss also decreases with increasing temperature as TLS are saturated by thermal phonons, making $Q_i$ increase up to $T\approx 1$ K, as shown in Fig. \ref{fig:resonator loss}(b). 
As the temperature becomes higher, $Q_i$ starts decreasing because of loss caused by thermally activated quasiparticles. The additional loss contribution as a function of temperature is
\begin{align}
\delta_{\mathrm{qp}} = \frac{\pi}{4} \frac{e^{\Delta_0/k_B T}}{\sinh(\hbar \omega/2 k_B T)K_0(\hbar \omega/2 k_B T)},
\end{align}
where $\Delta_0 = 1.76 k_B T_c$ is the superconducting gap of the NbTiN film and $K_0$ is the elliptic integral of the first kind \cite{zmuidzinasSuperconductingMicroresonatorsPhysics2012}. The total internal quality factor of the resonator is therefore
\begin{align}
\label{eq:low power loss model}
1/Q_i = 1/Q_{\mathrm{other}} + 1/Q_{\mathrm{TLS}} + 1/Q_{\mathrm{qp}},
\end{align}
where $1/Q_{\mathrm{other}}$ is the loss due to sources other than TLS and quasiparticles, and $Q_{\mathrm{qp}} = 1/\delta_{\mathrm{qp}}$. 
We find $T_c\approx$ 10 K and $Q_{\mathrm{other}} \sim 18000$ by fitting the data at $P_{\mathrm{in}}$ = -141 dBm (Fig. \ref{fig:resonator loss}(b)).

We also observe a temperature dependent frequency shift of the resonator $\delta f/f_0 = [f(T) - f(T=0)]/f(T=0)$ consistent with the presence quasiparticles at high temperatures as shown in Fig. \ref{fig:resonator loss}(c).
An expression for the temperature dependence of the frequency shift can be obtained from the complex conductivity of the superconducting film $\delta f/f_0 = (\alpha/2) \delta \sigma_2/\sigma_2$, where $\alpha$ is the kinetic inductance fraction and $\sigma_2$ is the imaginary part of the complex conductivity.
For low frequencies ($\hbar \omega \ll 2\Delta$), the shift can be written as \cite{tinkhamIntroductionSuperconductivity2004}
\begin{align}
\label{eq: qp freq shift}
\frac{\delta f}{f_0} = \frac{\alpha}{2}[\tanh(\Delta/2k_BT) - 1].
\end{align}
The solid line in Fig. \ref{fig:resonator loss}(c) is a fit to eq. \ref{eq: qp freq shift} where we assume that the kinetic inductance fraction $\alpha \approx 1$ and $\Delta \approx \Delta_0 = 1.76k_BT_c$ for the temperature range investigated here. The inset in Fig. \ref{fig:resonator loss}(c), shows a small frequency shift in the same temperature range where TLS contribute significantly to microwave losses. 
However, the frequency shift due to TLS is expected to be negative followed by a positive shift \cite{gaoPhysicsSuperconductingMicrowave2008} 
\begin{align}
\label{eq: TLS freq shift}
\frac{\delta f}{f_0} = \frac{\delta^0_{\mathrm{TLS}}}{\pi}[\mathrm{Re} \Psi (\frac{1}{2} + \frac{1}{2 \pi i} \frac{\hbar \omega}{k_BT}) - \ln(\frac{1}{2\pi}\frac{\hbar \omega}{k_BT})].
\end{align}
The initial decrease predicted by equation \ref{eq: TLS freq shift} using the parameter $\delta^0_{\mathrm{TLS}}$ of best fit from 10 mK data in Fig. \ref{fig:resonator loss}(a) is smaller than the error bars of our data, and the positive shift is smaller than that shown in Fig.\ref{fig:resonator loss}(c) inset by a factor of 10. Similar behavior has been observed in other resonators made from disordered films \cite{shearrowAtomicLayerDeposition2018}. 

\section{Tip-Sample Distance Calibration}

\begin{figure}
\includegraphics{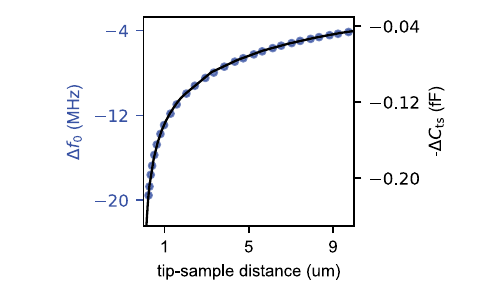}
\caption{\label{fig:tip sample distance} Measured approach curve (data points) show a decreasing resonance frequency as the distance between the tip and sample decreases. Solid black line is the simulated tip-sample capacitance $C_{\mathrm{ts}}$.}
\end{figure}

We estimated the distance between the tip and the sample by comparing the shift in the resonator frequency ($\Delta f_0$) with the shift in the tip-sample capacitance ($\Delta C_{\mathrm{ts}}$) relative to their values with the tip more than 100 $\mu m$ away from the sample. At such large tip-sample distances ($d_{\mathrm{ts}}$), the tip-sample interaction can be ignored and $f_0$ is not sensitive to small chance in $d_{\mathrm{ts}}$. In practice, the tip-sample distance is controlled by adjusting the displacement of the Z-scanner ($z$) under the sample stage $d_{\mathrm{ts}} = -z + d_0$, where $d_0$ is the tip-sample distance when the z scanner is at zero displacement. To determine $d_0$, we measured $\Delta f_0$ as a function of $z$ and simulated $C_{\mathrm{ts}}$ as a function of $d_{\mathrm{ts}}$ using the AC Conduction solver of Ansys Maxwell. When $C_{\mathrm{ts}}$ is small compared to the native capacitance of the resonator, they have the simple relation $\Delta C_{\mathrm{ts}} = \alpha \Delta f_0$, with the proportionality factor $\alpha$ to be a constant. We therefore find the values of $d_0$ and $\alpha$ for best fitting the two curves and quantitatively determine the $d_{\mathrm{ts}}$ dependence of $\Delta f_0$ (Fig. \ref{fig:tip sample distance}),
which allows us to estimate the distance between the tip and the sample using the measured frequency shift. For the experiments mentioned in the main text, we find $\alpha = 1.12\times 10^{-8} \mathrm{fF/Hz}$.

\begin{figure}
\includegraphics{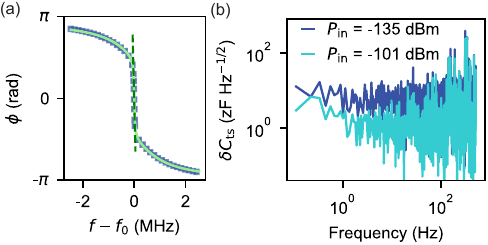}
\caption{\label{fig: sensitivity} Capacitive sensitivity. (a) A fit (solid line) to the measured phase response of the resonator (squares) can be used to convert phase noise into frequency noise using the slope of the fit (dashed line). (b) The tip-sample capacitance noise $\delta C_{\mathrm{ts}}$ for measurement bandwidth up to 500 Hz for both the few-photon regime (dark blue; $P_{\mathrm{in}} = -135 $ dBm) and higher powers (light blue; $P_{\mathrm{in}} = -101$ dBm). The data was taken with the tip at 500 nm above the Ta surface.}
\end{figure}

\section{Capacitance Measurement Sensitivity}
An expression for the sensitivity can be obtained from considering the frequency noise from the mechanical vibrations in the dilution refrigerator 

\begin{align}
\label{eq: frequency noise}
\delta f = S_{\phi}\frac{\partial{f}}{\partial{\phi}},
\end{align}

\noindent where $S_{\phi}$ is the amplitude noise spectral density of the phase ($\phi \equiv \angle S_{11}$) due to noise at our chosen tip-sample distance, and $\partial{f}/\partial{\phi}$ can be obtained from the slope of the response at the resonance frequency as shown in Fig. \ref{fig: sensitivity}(a) . 
This can be converted to an effective tip sample capacitance noise $\delta C_{\mathrm{ts}}$, using the fit parameter $\alpha$ from section V:
\begin{align}
\label{eq: capacitance noise}
\delta C_{\mathrm{ts}}  = \alpha \delta f.
\end{align}
To estimate our sensitivity at each measurement bandwidth, we computed $S_{\phi}$ using the Fourier transform of a time sequence of resonator phase signal measured at $f_0$. 
Using equations \ref{eq: frequency noise} and \ref{eq: capacitance noise}, we estimated the capacitive sensitivity of our sensor at each measurement bandwidth in Fig. \ref{fig: sensitivity}(b) at high power (light blue) and few photon (dark blue) operating conditions. 
In both cases, we achieve sensitivities on the order of zF/Hz$^{1/2}$. Higher power or longer measurement times would result in better sensitivities.  

\section{Transmon Loss Analysis}

\begin{figure*}
\includegraphics{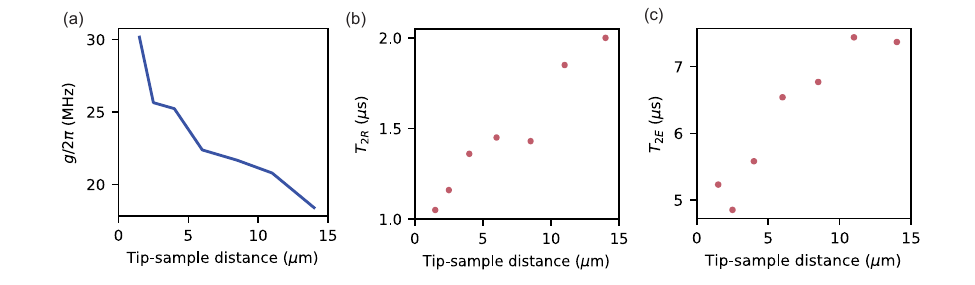}
\caption{\label{fig:transmon loss analysis} Tip-sample distance dependence of qubit-resonator coupling and qubit coherence. (a) Simulated qubit-resonator coupling rate $g/2\pi$ (b) Measured $T_{2R}$ and (c) Measured $T_{2E}$ at the same tip-sample distances as in Fig. 4.}
\end{figure*}
\begin{figure*}
\includegraphics{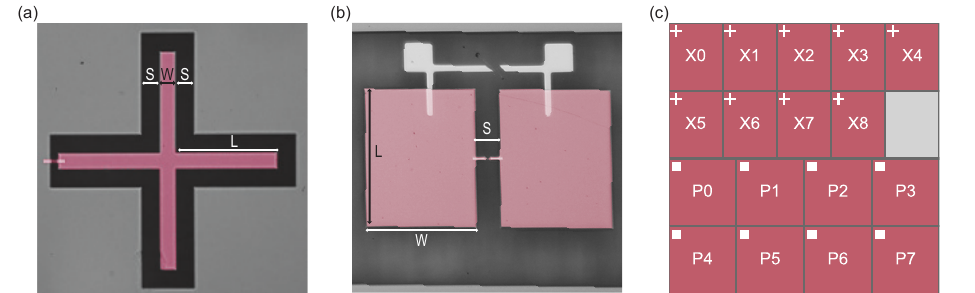}
\caption{\label{fig:transmon geometry} Transmon geometries and locations. (a) False-colored optical image of an Xmon qubit showing the critical geometrical parameters. $W$ and $L$ are the width and length of the arm of the center pad. $S$ is the spacing between the center pad and the ground plane. (b) False-colored optical image of a planar-capacitor transmon showing the critical geometrical parameters. $W$ and $L$ are the width and length of each capacitor pad. $S$ is the spacing between the capacitor pads. (c) Schematic showing the location of each transmon on the chip.}
\end{figure*}

\begin{table*}
\caption{\label{tab:table3} Summary of geometries,  characteristic energies, and Ramsey dephasing times ($T_{2R}$) of the transmons in Fig. 5. Devices X0-8 (P0-7) are Xmons (planar-capacitor transmons) with the critical dimensions $W,$ $L$ and $S$ defined as in Fig. \ref{fig:transmon geometry}. The charging energies ($E_C$) and the junction energies ($E_J$) are extracted using measured qubit frequencies and simulation using the AC Conduction solver of Ansys Maxwell and scQubits.}
\begin{ruledtabular}
\begin{tabular}{cccccccc}
Device&$W$ $(\mu \mathrm{m})$&$L$ $(\mu \mathrm{m})$&$S$ $(\mu \mathrm{m})$&$E_C$ (GHz)&$E_J$ (GHz)&$T_{2R}$ $(\mu \mathrm{s})$& \\ \hline
 X0&$20$&$94$ &$10$&$0.27$&$22$&$1.8$\\
 X1&$30$&$94$&$15$&$0.28$&$21$&$1.6$\\
 X2&$30$&$118$&$30$&$0.29$&$20$&$1.7$\\
 X3&$15$&$118$&$15$&$0.27$&$23$&$2.4$\\
 X4&$20$&$118$&$20$&$0.28$&$23$&$2.2$\\
 X5&$30$&$225$&$30$&$0.16$&$42$&$1.3$\\
 X6&$30$&$105$&$30$&$0.31$&$26$&$0.8$\\
 X7&$30$&$100$&$30$&$0.33$&$20$&$1.1$\\
 X8&$30$&$118$&$30$&$0.29$&$18$&$1.8$\\
 P0&$200$&$280$&$40$&$0.27$&$32$&$0.7$\\
 P1&$200$&$250$&$40$&$0.30$&$27$&$0.9$\\
 P2&$200$&$280$&$40$&$0.27$&$25$&$1.7$\\
 P3&$200$&$330$&$40$&$0.25$&$28$&$1.0$\\
 P4&$200$&$310$&$80$&$0.25$&$28$&$1.0$\\
 P5&$200$&$235$&$10$&$0.24$&$27$&$1.6$\\
 P6&$200$&$280$&$40$&$0.27$&$29$&$1.4$\\
 P7&$200$&$265$&$20$&$0.25$&$29$&$1.1$\\
 
\end{tabular}
\end{ruledtabular}
\label{tab:transmon geometries and characterization results}
\end{table*}

To estimate the Purcell loss caused by the resonator, we simulated the qubit-resonator coupling rate ($g$) using the AC Conduction solver of Ansys Maxwell and scQubits \cite{groszkowskiScqubitsPythonPackage2021}. The Purcell loss rate ($\gamma_{Purcell}$) is then calculated by

\begin{align}
\label{eq: Purcell loss}
\gamma_{Purcell} = (\frac{g}{\Delta})^2\kappa
\end{align}
where $\Delta = \omega_q - \omega_r$ and $\kappa$ are the measured qubit-resonator detuning and resonator linewidth. Figure \ref{fig:transmon loss analysis}(a) shows the simulated $g/2\pi$ as a function of tip-sample distance for estimating $\Gamma_1$ in Fig. 4. We also measured $T_{2R}$ and $T_{2E}$ as the function of the tip-sample distance at the same location and the results are shown in Fig. \ref{fig:transmon loss analysis}(b),(c).

\section{Transmon Geometries and Characterization Results}
Two types of transmons were measured in this work: Xmons whose capacitor is formed by a four-arm center pad and the ground plane (Fig. \ref{fig:transmon geometry}(a)), and planar-capacitor transmons with two floated capacitor pads (Fig. \ref{fig:transmon geometry}(b)). The geometries of individual transmons are varied throughout the chip and are provided in Table \ref{tab:transmon geometries and characterization results}. Fig. \ref{fig:transmon geometry}(c) shows the locations of the transmons on the chip. The characteristic energies and Ramsey dephasing times of all characterized transmons are also given in Table \ref{tab:transmon geometries and characterization results}.

\bibliographystyle{aipnum4-1}
\bibliography{supp}